# Review on Biophysical Modelling and Simulation Studies for Transcranial Magnetic Stimulation

Jose Gomez-Tames[1], Ilkka Laakso[2], and Akimasa Hirata[1]

1. Nagoya Institute of Technology, Department of Electrical and Mechanical Engineering, Nagoya, Aichi 466-8555, Japan
2. Department of Electrical Engineering and Automation, Aalto University, FI-00076, Finland.

**Corresponding author:**

Jose Gomez-Tames

Nagoya Institute of Technology

Tel & Fax: +81-52-735-7916

E-mail:jgomez@nitech.ac.jp

*Abstract*

Transcranial magnetic stimulation (TMS) is a technique for noninvasively stimulating a brain area for therapeutic, rehabilitation treatments and neuroscience research. Despite our understanding of the physical principles and experimental developments pertaining to TMS, it is difficult to identify the exact brain target as the generated dosage exhibits a non-uniform distribution owing to the complicated and subject-dependent brain anatomy and the lack of biomarkers that can quantify the effects of TMS in most cortical areas. Computational dosimetry has progressed significantly and enables TMS assessment by computation of the induced electric field (the primary physical agent known to activate the brain neurons) in a digital representation of the human head. In this review, TMS dosimetry studies are summarised, clarifying the importance of the anatomical and human biophysical parameters and computational methods. This review shows that there is a high consensus on the importance of a detailed cortical folding representation and an accurate modelling of the surrounding cerebrospinal fluid. Recent studies have also enabled the prediction of individually optimised stimulation based on magnetic resonance imaging of the patient/subject and have attempted to understand the temporal effects of TMS at the cellular level by incorporating neural modelling. These efforts, together with the fast deployment of personalised TMS computations, will permit the adoption of TMS dosimetry as a standard procedure in clinical procedures.

# 1. Introduction

Transcranial magnetic stimulation (TMS) is a technique for noninvasively stimulating a target area of the brain. TMS is used for diagnosis in pre-surgical identification of motor and language functions and in the treatment of neurological diseases or conditions. Since the first study on TMS (Barker, Jalinous and Freeston 1985), this field has grown substantially. One difficulty is the lack of available biomarkers to investigate the effects of TMS in the various regions of the brain, excluding the somatosensory, visual, and language regions. Moreover, recent studies report that the electric field in the brain exhibits a non-uniform distribution owing to the complicated and subject-dependent brain anatomy (Thielscher, Opitz and Windhoff, 2011; Janssen *et al.*, 2013; Nummenmaa *et al.*, 2013; Laakso, Hirata and Ugawa, 2014), which results in a greater challenge while estimating the target regions. Further, the estimated regions by TMS may vary according to different parameters, such as the variations in the type of magnetic coil, its position and orientation over the scalp, and the current waveform injected into the coil (Taniguchi, Cedzich and Schramm, 1993; Terao and Ugawa, 2002; Holsheimer *et al.*, 2007). Thus, the necessity to understand, visualise, and individually optimise the TMS dosage in the brain has motivated the application of computational dosimetry.

For the past 30 years, computational dosimetry has progressed significantly in biophysical and electrophysiological modelling techniques to investigate the effects of electromagnetic fields in the human body. In TMS, the primary physical agent known to activate neurons is the induced electric field (EF); moreover, recent studies have enabled the prediction of the stimulation location and optimised the dosages of the stimulation parameters (Opitz *et al.*, 2014; Aonuma *et al.*, 2018; Seynaeve *et al.*, 2019; Weise *et al.*, 2020). These procedures estimate the EF in the brain while considering the effects of the various physical aspects involved in TMS (e.g. biological tissue conductivity, coil design, head anatomy). Recent studies considered the connection between induced EF and neuronal responses, which was based on a three-staged computation. The first step (subsection 2.1) involves expressing a human body as discrete geometric elements with millimetre or sub-millimetre resolution based on medical images. The second step (subsection 2.2) is the determination of the induced EF in the brain. The third step (subsection 2.3) involves modelling the neural responses evoked by TMS at the cellular level.

Reviews and guidelines for TMS were published from the clinical perspective *(Rossi et al.*, 2009; Perera *et al.*, 2016). However, no TMS review has been published on the simulation techniques for biophysical modelling to the best of our knowledge, despite the rapid increase in the number of computational studies. The current review presents the historical and most recent efforts on TMS modelling. This review is intended for research groups working on dosimetry for clinical applications and researchers working on the clinical and neuroscience aspects of TMS, who are interested in adopting computational models.

## 2. Outline of Computational Models

TMS-induced EF is computed by using a realistically segmented human head model that represents the tissue-dependent conductivity distribution. The general pipeline for TMS modelling in an individualised head model is illustrated in figures 1 and 2. Further, the EF effects on a neuronal model can be investigated by using a compartmentalised cable equation, as shown in figure 3. The outline of the implementation process is described in the following subsections.

### 2.1 Development of a Personalised Head Model

The construction of a human body model has progressed corresponding to the developments and improvements in medical images and their processing. In the 1990s and early 2000s, when image processing performance was inadequate, a certain amount of manual assessment was required for human tissue classification to construct human body models. The models were presented (Zubal *et al.*, 1994; Dimbylow, 1997; Nagaoka *et al.*, 2004) as voxel phantoms and subsequently expanded to 'families' or 'populations' of phantoms (Christ, Kainz and Hahn, 2010; Wu *et al.*, 2012; Park *et al.*, 2018). In the last 10 years, it has become possible to construct personalised head models almost automatically from magnetic resonance imaging (MRI) data, owing to progress in medical imaging techniques (Dale, Fischl and Sereno, 1999; Fischl, 2012; Windhoff, Opitz and Thielscher, 2013; Laakso *et al.*, 2015; Lee *et al.*, 2016; Huang *et al.*, 2019).

Automatic segmentation of the brain and non-brain tissues from MRI data can be obtained using different image analysis software, such as FreeSurfer (Dale, Fischl and Sereno, 1999; Fischl, 2012), Statistical Parametric Mapping (SPM) (Ashburner and Friston, 2005), and FMRIB software library (FSL) (Smith, 2002). These different image analysis tools have been incorporated into different head model generation and EF calculation pipelines, such as ROAST (Huang *et al.*, 2019) and SimNIBS (Windhoff, Opitz and Thielscher, 2013). An illustration of a pipeline to segment the brain and non-brain tissues is shown in figure 1 (Windhoff, Opitz and Thielscher, 2013; Laakso *et al.*, 2015; Huang *et al.*, 2019).

## 2.2 Electromagnetic Computation

Computational electromagnetic methods are based on the quasi-static approximation to determine induced EF at frequencies lower than several megahertz (Barchanski *et al.*, 2005; Hirata, Ito and Laakso, 2013), as shown in figure 2. In the quasi-static approximation, Maxwell's equations can be simplified by ignoring propagation, capacitive, and inductive effects (Plonsey and Heppner, 1967), which result in the following equation for the electric scalar potential:

$$\nabla \cdot \left[ \sigma \left( -\nabla \varphi - \frac{\partial}{\partial t} \boldsymbol{A}_0 \right) \right] = 0, \qquad (1)$$

where $\boldsymbol{A}_0$ and $\sigma$ denote the magnetic vector potential of the applied magnetic field and the tissue conductivity, respectively. If the induced current marginally perturbs the external magnetic field, then $\boldsymbol{A}_0$ is equal to the magneto-static vector potential that is completely decoupled from the EF and can be calculated by considering the Biot–Savart law pertaining to the source current distribution. At the boundary of the body, the scalar potential satisfies the Neumann boundary condition:

$$\mathbf{n} \cdot \nabla \varphi = -\mathbf{n} \cdot \frac{\partial}{\partial t} \boldsymbol{A}_0, \qquad (2)$$

where **n** is a normal vector to the body surface.

Once equation (1) is solved, the induced EF **E** can be expressed as:

$$\boldsymbol{E} = -\nabla \varphi - \frac{\partial}{\partial t} \boldsymbol{A_0}. \qquad (3)$$

The factor $\partial \boldsymbol{A}_0 / \partial t$ is called the primary EF, which is induced by the changing magnetic field, which depends only on the TMS coil characteristics. The factor $-\nabla \varphi$ is called the secondary field, which is caused by charges in the conducting medium. The induced current density and EF can be related in terms of $\boldsymbol{J} = \sigma \boldsymbol{E}.$ Equation (1) typically has no analytical solution. Instead, numerical methods must be used to approximate $\varphi$. Several computational electromagnetic methods can be applied to solve equation (1) including finite element method (FEM), boundary element method (BEM), and finite-difference method (FDM).

In the FEM, the geometry of the computation domain, in this case, the head, is divided into a mesh of several small non-overlapping finite elements, typically tetrahedrons or hexahedrons. Then, the mesh is used to define a set of basis functions, which are non-zero only in a small number of elements and are polynomials of a specified order inside each element. The approximate solution to equation (1) is calculated as a linear combination of these basis functions by solving a large linear equation system.

BEMs are derived by converting equation (1) to an integral equation form. The integral equation can be formulated using either the induced scalar potential (e.g. Ferguson and Stroink, 1997) or the induced surface charge (e.g. Makarov *et al.*, 2018) as the unknown. If the geometry consists of a finite number of compartments, each having a uniform conductivity, the problem reduces to solving the unknown surface potential or charge on each interface between the compartments. In TMS literature, the tissue interfaces are commonly referred to as 'layers'. The integral equation for the surface potential/charge is solved numerically using the FEM. Subsequently, the surface potential/charge can be used to calculate the induced potential and EF at any location.

FDM formulations are derived by replacing the spatial derivatives in equation (1) by their finite-difference approximations, which results in a linear equation system from which the unknown potential values at each element can be obtained. FDM resembles the FEM in the special case of a structured mesh consisting of hexahedral elements. An example of FDM is the scalar potential finite difference (SPFD) method (Dawson and Stuchly, 1996), which uses the second-order central difference approximation to the left side of equation (1).

Each method has advantages and disadvantages when it comes to modelling TMS. FEM and BEM can use unstructured computational meshes that can accurately represent curved boundaries between tissues. The mesh can furthermore be locally refined near the targeted brain areas, thereby improving accuracy (Windhoff, Opitz and Thielscher, 2013). The weakness of unstructured meshes is that the generation of a good quality mesh from the segmented MRI data is a non-trivial task. In contrast, FDM is limited to a structured rectangular grid, which is trivially obtained from the segmented images, but results in 'staircase' approximation of curved boundaries. FEM approaches that use structured grids also suffer from the staircase approximation error. BEM differs from FEM and FDM as it only requires the meshing of the boundaries between the tissue compartments. Each compartment must have a uniform conductivity; consequently, the method cannot efficiently model anisotropic or heterogeneous materials, which can be modelled using FEM or FDM in a straightforward manner (Wang and Eisenberg, 1994; Windhoff, Opitz and Thielscher, 2013).

Gomez *et al.* (2020) compared the accuracy of FEM, BEM, and FDM for modelling TMS-induced EFs in a realistic head model. Using computational meshes of varying resolutions as well as basis functions of different polynomial orders, they showed that all three methods could produce accurate results, provided that the resolution of the computational mesh was sufficiently fine. To obtain a desired level of numerical accuracy, the required mesh resolution depended on the method and the order of the elements.

Saturnino *et. al.* (2019) and Soldati and Laakso (2020) obtained similar results, showing how the error in the EFs calculated using the FEM diminished when the resolution of the mesh (tetrahedral or cubical elements) was refined. These findings indicate that the numerical errors can be controlled. Therefore, all computational methods, if their parameters have been set appropriately, can produce sufficient computational accuracy for TMS modelling studies.

**2.3 Multiscale Model Incorporating Neural Modelling**

A cable equation is used to describe propagation and interaction of electrical signals in the axons of neurons, as shown in figure 3. Further, it can incorporate the ionic mechanisms underlying the initiation and propagation of action potentials. Hodgkin and Huxley proposed the first model for neural signal propagation in a squid giant axon (Hodgkin and Huxley, 1952). The seminal study resulted in the development of subsequent models of excitable cell membrane (Frankenhaeuser and Huxley, 1964; Chiu and Ritchie, 1979; Sweeney, Mortimer and Durand, 1987; McIntyre, Richardson and Grill, 2002) including the brain (Aberra, Peterchev and Grill, 2018). The original cable equation was initially modified to include responses to electric and magnetic stimulation (McNeal, 1976; Rattay, 1986; Reilly, 1989; Roth and Basser, 1990); this allowed examinations of the spinal cord, muscle, and brain stimulation using realistic models (Doheny *et al.*, 2008; Wongsarnpigoon and Grill, 2008; Danner *et al.*, 2011; Salvador *et al.*, 2011; Seo, Kim and Jun, 2015; Aberra *et al.*, 2020).

A modified cable equation describes the neuronal membrane polarisation and activation due to TMS-induced EF

$$c_m \frac{dV_m}{dt} = -I_{ion} + \frac{1}{R}\frac{d^2 V_m}{ds^2} + \frac{1}{R}\frac{d^2 V_e}{ds^2}, \qquad (4)$$

where $c_m$, $R$, and $I_{ion}$ denote the membrane capacitance, the intra-axonal resistance, and the ionic current of the membrane per unit length, respectively. The spatial variable *s* is the distance along the trajectory of the neuron. The term $V_m$ denotes the membrane potential along the cable, the quasi-potential $V_e$ is the line integral of the induced EF

$$V_e(s,t) = -\int_0^S \boldsymbol{E}(\boldsymbol{r}(s'),t) \cdot d\boldsymbol{s}', \tag{5}$$

where $\boldsymbol{r}$ is the arc length parametrisation of the path of the neuron.

The cable equation (4) can be modelled in the compartmental form so that different sections of the neuron are approximated by an electric network. Each compartment $n$ consists of axial resistance, membrane conductance, and capacitance. The membrane potential in each compartment can be determined from

$$c_{m,n}\frac{dV_{m,n}(t)}{dt} = -I_{ion,n} + \frac{V_{m,n-1}(t) - 2V_{m,n}(t) + V_{m,n+1}(t)}{R} - \frac{V_{e,n-1}(t) - 2V_{e,n}(t) + V_e(t)}{R} \tag{6}$$

Finally, it is to be noted that the electrical properties of each compartment depend on the neuron segments, so the model can be extended to include morphologically detailed neurons for the central nervous system (CNS) (Aberra, Peterchev and Grill, 2018). The model can consist of bifurcations and branches, pre- and post-synaptic terminals, and dendritic arborisation.

## 3. Electric Field Dosimetry

In the early 1990s, it was possible to calculate the EF strength using spherical models of the brain (Cohen *et al.*, 1990; Tofts, 1990; Eaton, 1992). However, the location and extent of the EFs are affected by anatomical factors (e.g., gyrification) and the electrical conductivities of different tissue types, which can only be accounted for using anatomical models. The progress of TMS dosimetry is summarized in figure 4 that shows that transition between simplified to anatomical models. Simplified head models are listed in table 1, and complexity variation of the anatomical head model and its targets (population segments and brain regions) are summarised in figure 5.

In this section, subsections 3.1 – 3.4 deal with fundamental aspects of TMS modelling, such as modelling of the anatomy, electrical conductivity and magnetic coils, whereas subsections 3.5 – 3.8 deal with TMS dosimetry applications, such as coil design optimisation, guiding TMS dose, and comparison of EF dosimetry with experimental measurements. The identified studies in this review are based on a search strategy presented in Appendix A that was developed for each subsection. Finally, the data used in figure 5 was retrieved from the identified studies in subsections 3.1 and 3.5 – 3.8. In the case of Table 1, the data was retrieved from studies in subsections 3.1 and 3.3 – 3.6.

### 3.1 Representation of head tissues

The initial attempts to compute the TMS-induced EF used the brain representations as an infinite half-space or as spheres. However, as illustrated in figure 6, the lack of anatomical detail limits the accuracy of the estimated EF. The following seven studies that investigated the effects of tissue representation on the induced EF were identified.

Wagner *et al.* (2004) considered a five-layered model to investigate the effects of tissue inhomogeneity and geometry on the induced current density using the FEM. Their study showed that the boundaries between tissues of different conductivities strongly affected the distribution of the current density. In particular, inhomogeneity of the brain produced a normal component of the current density, which is absent in spherical head models and which was equivalent to 30% of the total current density at the target region.

Toschi *et al.* (2008) used a realistic inhomogeneous head model, which was derived from MRI data, and computed the induced EF using FDM. They demonstrated that TMS with a symmetrical distribution of the primary field, such as in a figure-of-eight coil, induces a highly asymmetrical EF distribution in a realistic anatomy. The authors concluded that a high-resolution field solver and a realistic reconstruction of the head geometry of the subject are required for a highly accurate prediction of the induced EF.

Salinas, Lancaster and Fox (2009) used a realistic six-layer head model to assess the effect of multiple layers on the EF strength using the BEM. The secondary EF is important as its strength ranges from 20 to 35% of that of the primary EF. The authors concluded that an accurate tissue geometry representation is required to consider the secondary EF effects accurately.

Silva, Basser and Miranda (2008) considered tissue heterogeneity in a layered cortical sulcus model to investigate the spatial distribution of the induced EF. The primary finding was that the electrical conductivity and cortical folding should be considered to estimate stimulation regions.

Thielscher, Opitz and Windhoff (2011) used the FEM to characterise the induced EF in a head model that considered the realistic gyrification patterns. Five tissue types were considered. The induced EF strength in grey matter was increased by up to 50% when the induced current was perpendicular to the local gyral orientation in comparison with a simplified homogeneous model that neglected cortical gyrification. In contrast, the EF direction was predominantly influenced by the CSF-skull boundary. In general, when compared to the anatomical model, the spherical head model presented lower maximal field strengths and a lower focality of the field in grey matter and did not show spatial shifts corresponding to the TMS coil orientation changes.

Nummenmaa *et al.* (2013) compared different head models with different levels of detail: spherical, semi-anatomical (skin, skull, and intracranial without CSF or gyrification), and anatomical (skin, skull, CSF, and brain) using the BEM. The results showed that anatomical and semi-anatomical head models demonstrated similar induced EF distributions, although the former had higher EF strength. In contrast, the spherical model did not reproduce distribution similar to the anatomical model.

Janssen, Oostendorp and Stegeman (2014) incorporated geometrical detail, specifically for a highly detailed CSF-grey matter boundary in an FEM model. They concluded that omitting the secondary field due to charge accumulation at the boundaries of the tissue significantly affects the total induced EF distribution and strength.

In summary, inhomogeneous anatomical head models are required for the accurate estimation of the induced EF (Wagner *et al.*, 2004; Silva, Basser and Miranda, 2008; Thielscher, Opitz and Windhoff, 2011; Janssen, Oostendorp and Stegeman, 2014; Bungert *et al.*, 2017). Homogeneous models cannot describe the effects of coil orientation dependency, and the secondary field effects are omitted; consequently, the induced EF distribution is not accurately predicted. The boundaries between tissues of high contrast conductivities can strongly affect the induced EF distribution (Toschi *et al.*, 2008; Salinas, Lancaster and Fox, 2009; Nummenmaa *et al.*, 2013; Janssen, Oostendorp and Stegeman, 2014; Bungert *et al.*, 2017).

### 3.2 Electrical conductivity: variability

The selection of the electrical conductivity of the tissues is a challenge; however, it is sometimes a controversial topic owing to the lack of defined values and diversity of reported values (Saturnino, Madsen and Thielscher, 2019). Their selection becomes important as high contrast conductivity between neighbouring tissues significantly affects induced EF distribution, as discussed in the previous subsection. Here, we initially reviewed the variability of electrical properties of the tissues used in TMS modelling studies. In total, 66 modelling studies were identified; the reported conductivities values are summarized in Figure 7.

The choice of the conductivity values is predominantly based on the conductivity values selected by Wagner *et al.*, (2004) or the tissue dielectric property database presented by Gabriel, Lau and Gabriel (1996) at frequencies similar to the TMS operating frequency (2.5 to 10 kHz). The conductivity of CSF is relatively constant across different studies (1.65 to 2.0 S/m). The grey matter and white matter have relatively small variations, i.e. between 0.1 to 0.276 S/m and 0.07 to 0.126 S/m, respectively. These differences may not be significant in the localisation of the highest EF strengths in the cerebral cortex, according to (Aonuma *et al.*, 2018; Gomez-Tames *et al.*, 2018).

Scalp and skull present larger variabilities. The variability of the former is between 0.0002 to 0.465 S/m, and the latter between 0.001 to 0.08 S/m. In the case of the scalp, the innermost and outermost layers of the skin present large differences. The lower bound (0.0002 S/m) may be related to the outermost stratum corneum layer of the skin (Yamamoto and Yamamoto, 1976). The upper bound (0.465 S/m) is based on the measurements at direct current (Burger and Milaan, 1943). Certain studies adopted an average value between the fat (0.02 to 0.08 S/m) and muscle (0.2 to 0.4 S/m) when they considered them together as the scalp layer, such as (Bungert *et al.*, 2017) or non-uniform values within the same tissue (Rashed, Gomez-Tames and Hirata, 2020a). Although these non-brain tissues (except the CSF) do not significantly affect the effects of the TMS on the brain tissues (Saturnino, Madsen and Thielscher, 2019), non-brain tissues are required to investigate the side-effects and safety.

### 3.3 Electrical conductivity: anisotropy

Anisotropic electrical characteristics of the brain, in particular the white matter, may affect the modelled EF. In the white matter, owing to the presence of interconnecting neural tracts, the conductivity in directions along and across the neural tracts may differ by a factor of ten (Nicholson, 1965). Four studies were identified that investigated the effect of anisotropy on the induced EF.

Miranda, Hallett and Basser (2003) showed a significant difference in the induced EF when considering anisotropic conductivity in an inhomogeneous spherical model using the FEM.

De Lucia *et al*. (2007) used a realistic head model that considered anisotropic conductivity derived from diffusion tensor imaging in the brain. The induced EF in the part of the grey matter was marginally affected by the tissue anisotropy in an FEM model. Instead, the induced EF strength variations were approximately 10% between models using isotropic or anisotropic conductivity in the white matter.

Opitz *et al.* (2011) showed that considering an anisotropic brain tends to enhance the local EF hotspots in white matter by 40% in an FEM model; however, no changes were observed in the grey matter.

De Geeter *et al.* (2012) investigated the effect of realistic dispersive anisotropic tissue properties on the induced EF in a head model with realistic geometry using an FDM (impedance method). The results showed that anisotropy yields a difference of up to 19% on the maximum EF in the white matter (a mid-value between two previous studies (De Lucia *et al.*, 2007; Opitz *et al.*, 2011)), while the differences in the other tissues were not significant.

### 3.4 TMS coil models and verification

The accuracy of the TMS-induced EF depends on the level of detail of the magnetic coil. Three primary approaches have been used: modelling the coil as a collection of thin wires (Eaton, 1992), magnetic dipoles (Ravazzani *et al.*, 1996), or realistic models that consider the current distribution in the coil windings (Salinas, Lancaster and Fox, 2007). Verification of the correct modelling can be conducted by direct measurements of the induced EF in experimental phantoms. Seven studies that have considered these approaches for TMS coil modelling have been identified.

Thielscher and Kammer (2004) computed the induced EF by the superposition of the fields of appropriately placed magnetic dipoles using an X-ray for modelling the coils by extending the method presented by Ravazzani *et al.* (1996).

Salinas, Lancaster and Fox (2007) presented a detailed TMS coil wiring geometry, which considered the width, height, shape, and number of turns of the wire. The induced EF computed using a detailed TMS coil model had an error within 0.5% with respect to the measurement values. A realistic approximation of the TMS coil model is more important near the coil than at the cortical depth, where no significant difference was measured between simple (thin wire) and detailed coil models.

Tachas, Efthimiadis and Samaras (2013) modelled a figure-of-eight coil with different degrees of accuracy to investigate the impact on the induced EF distribution in a realistic head model using an FDM (impedance method). Modelling the figure-of-eight coil using two single thin-wire loops yielded inaccurate induced EF distributions. Double thin-wire loops compared well to more realistic spiral-based approach.

Petrov *et al.* (2017) evaluated different models of a figure-of-eight TMS coil with different levels of modelling complexity: simple circular coil model, coil with in-plane spiral winding turns, and coil with stacked spiral winding turns. The thickness of the coil winding affected the induced EF minimally. However, modelling the in-plane coil geometry was important to simulate the induced EF accurately and to ensure reliable predictions of neuronal activation.

Nieminen, Koponen and Ilmoniemi (2015) introduced an instrument for automated measurement of the E-fields induced by TMS coils in spherically symmetric conductors approximating the head. Later, Çan *et al.* (2018) modelled three types of TMS coils using thin-wire approximation for current loops. The calculations and measurements in a spherical phantom showed that the induced EF distribution was highly consistent with the measurements for all coil types.

Gomez *et al.* (2020) showed that magnetic coil models constructed from magnetic or current dipoles produced errors smaller than 2% in the primary EF when compared to thick solid-conductor coils. The error could be further reduced by increasing the number of dipoles. Ignoring the eddy currents in the coil windings could generate a maximum point-wise error below 5% of the induced EF in a spherical model.

In summary, the modelling fidelity of the TMS coil was revised in different studies. It was determined that single thin-wire loops representing the coil were inaccurate (Tachas, Efthimiadis and Samaras, 2013; Petrov *et al.*, 2017). If the winding arrangement significantly resemble the experimental coil, the induced EF was consistent with experimental measurements (Çan *et al.*, 2018). Overall, the methods typically used for modelling magnetic coils can sufficiently suppress numerical errors (Gomez *et al.*, 2020). Finally, the comparison between computed and measured induced EFs demonstrated good agreement, suggesting good confidence in the dosimetry techniques (Salinas, Lancaster and Fox, 2009; Nieminen, Koponen and Ilmoniemi, 2015; Çan *et al.*, 2018).

### 3.5 Effects of anatomical and inter-individual factors

Adopting realistic head models is a requirement to achieve good accuracy of the computed induced EF when compared to simplified geometries (subsection 3.1). Inter-subject variability and specific anatomical aspects affect the induced EF. The extent of inter-individual variability affects how well the findings obtained in one head model can be generalised to a population. Here, we review a total of ten papers accounting for the effects of individual anatomical factors on the induced EF.

Opitz *et al.* (2011) showed that the induced EF strength depends on the individual cortical folding pattern using the FEM. The EF strength is selectively enhanced at the gyral crowns and lips, and high EF strength can also occur deep in the white matter. These effects might create hot spots in white matter, resulting in potential neural excitation.

Bijsterbosch *et al.* (2012) demonstrated that subject-specific gyral folding patterns and local thickness of subarachnoid CSF are necessary to determine potential stimulation sites accurately using the FEM. Their computation showed that high induced EFs occurred primarily on the crowns of the gyri which had only a thin layer of CSF above them. Consequently, the peak EFs can occur in grey matter regions distant from the assumed spot underneath the centre of the figure-of-eight coil depending on the local variations of CSF thickness. Further, the authors compared two subjects (male and female). The female model had a lower peak intensity (0.6 times lower), partially owing to the larger scalp-cortex distance.

Janssen *et al.* (2013) investigated the effect of the sulcus width (< 1.5 mm) on the induced EF strength in a head model using the FEM. They determined that the sulcus width did not cause large differences in the majority of the EF strengths. However, considerable overestimation of sulcus width (and consequently thin gyri) produced an overestimation of the calculated EF strength, which also occurred at locations distant from the target location.

Opitz *et al.* (2013) generated realistic head models of five subjects and used the FEM to compute the induced EF distribution on the motor cortex. The authors observed that individuals having a hand motor cortex that was shaped like an inverted omega responded preferentially to a 45° coil orientation, while one subject having a hand motor cortex shaped like an epsilon responded preferentially to a 90° coil orientation.

Crowther, Hadimani and Jiles (2014) showed a significant difference in the induced EF between four models (adult man, adult woman, girl, and boy). Higher EF strength was observed in younger and smaller brain models.

In Yamamoto *et al.* (2016), six individual head models were constructed by segmenting MRI data. The SPFD method was used to compute the induced EF strength at resting motor threshold (RMT) in the motor cortex of each subject. The EF strengths on the target region had a normalised standard deviation of 18% (mean value of 203 V/m).

Lee *et al.* (2016) investigated how the induced EF is affected by brain-scalp distance using heterogeneous head models constructed from MRI data of 50 subjects (with a maximum age of 36 years). With an increment in brain-scalp distance, the maximum EF decreased while the stimulation area increased.

Laakso *et al.* (2018) calculated the induced EF strength in 19 subjects using the FEM. The maximum EF strength calculated at active motor threshold (AMT) and RMT had normalised standard deviations of 19% and 15% (mean values of 129 V/m and 166 V/m), respectively. The same group (Can *et al.*, 2019) extended the analysis to cerebellar TMS for the same subjects. The normalised standard deviations of the maximum EF strength in the cerebellum ranged between 10 to 20%, depending on the type of the magnetic coil and its location.

Gomez-Tames *et al.* (2018) determined for each point in the cortex the coil location and orientation that maximized the induced EF strength using the SPFD method. Between 18 subjects, the normalised standard deviation of the maximum EF strength varied from 5 to 40%, with an average of 20%. The variability of the maximum EF strength was minor at the motor or sensory areas where the sulcus was approximately in the same direction in all individuals. The variability was larger in other regions, which had complicated, variable, and distinct folding patterns between individuals.

Zhong *et al.* (2019) demonstrated the difference between two coils (conventional figure-of-eight coil and the coil used for deep brain stimulation) targeting the cerebellum in 50 subjects using the FEM. The maximum induced EF strength had a normalised standard deviation ranging from 20 to 34% among subjects in the target regions.

In summary, various studies have investigated inter-subject variability of the induced EF ranging from a few to up to 50 subjects (Bijsterbosch *et al.*, 2012; Crowther, Hadimani and Jiles, 2014; Lee *et al.*, 2016; Yamamoto *et al.*, 2016). In the studies reviewed in this and other subsections, adults have been the predominant population segment (figure 5(b)), and the elderly and youth populations are almost unexplored. Further, TMS targeting the prefrontal and motor-sensory areas accounted for 74% of the studies, followed by deep and cerebellar areas (16%). Parietal, temporal, and occipital accounted for 10% of the studies, as illustrated in figure 5(c).

### 3.6 Coil design: optimisation and performance

The circular coil was the first design used for TMS in the seminal work presented in (Barker, Jalinous and Freeston, 1985). The first successful attempt to optimise the TMS coil for the focality improvement used the figure-of-eight coil, which was presented by (Ueno, Tashiro and Harada, 1988), and others coil have also been adopted in clinical research, as illustrated in figure 8. Here, we identified 23 studies that have used EF calculations to investigate coil design, optimization, and performance. As listed in table 2, the studies have used various metrics, such as depth and spread of the induced EF or energy requirements, to optimise and study the performance of various coil designs.

Roth, Zangen and Hallett (2002) proposed the first coil designed for the stimulation of deep brain regions termed Hesed coil (H-coil). The H-coil demonstrates a slower decrease of the induced EF as a function of the distance from the coil centre than that in double cone and circular coils. This was confirmed by phantom measurements and numerical computation.

To reduce the spread of the induced EF and improve focality, Kim, Georghiou and Won (2006) computed the effect of passive shielding plates that partially blocked the electric and magnetic fields. One disadvantage of this method was a reduction in the maximum EF when using the shield plate. A reduction of 50% in the maximum EF strength was observed at a distance of 40 mm from the coil.

Im and Lee (2006) evaluated a multi-channel TMS system using realistic simulations using up to 128 small coils. Using this system, enhanced targeting accuracy and concentrated induced EF distribution was possible.

Lu *et al.* (2009) presented a multi-channel TMS system with 40 small coils. The induced current density and EF in a realistic human head model were calculated using the FDM. Proper adjustment of the input current phases can improve the induced EF strength in the brain, although coil size does not allow strong fields, such as in the figure-of-eight coil.

Salvador *et al*. (2009) showed that a high permeability core in an H-coil could increase focality and field intensity by 25%. The performance of the proposed design was investigated using a realistically shaped homogeneous head model.

Hernandez-Garcia *et al*. (2010) considered active shielding of the TMS coil by using a secondary coil, which created opposing electric and magnetic fields that cancelled the field of the source outside the region of interest. Iterative optimisation techniques were used to design active shields for the figure-of-eight coil by considering two objectives: selectivity and depth of the primary EF computation for a spherical model. The resulting designs were tested on a realistic human head model. For the same penetration depth between shielded and unshielded cases, the volume was reduced by 13% for the shielded case relative to the unshielded case.

Deng, Lisanby and Peterchev (2013) quantified the spread and depth of the induced EF to characterise the performance of 50 TMS coils using a spherical model. For any coil design, the ability to directly stimulate deeper brain structures was obtained at the expense of inducing a wider electrical field spread; moreover, none of the coil designs was able to overcome the depth–focality trade-off. However, the figure-of-eight-shaped coils were more focal (the area where the field strength becomes half of the maximum; 5 cm$^2$) when compared to circular coils (34 cm$^2$).

Deng, Lisanby and Peterchev (2014) showed that larger coils were more appropriate for deep TMS by analysing the depth–focality trade-off of the EF in a spherical head model. Coils with larger diameters had an EF that decays slower in depth but was less focal than that of smaller diameters. Although smaller coils had superior focality than the larger coil, the advantage in terms of activated brain volume diminished with increasing target depth. The double cone coil offers high energy efficiency and balance between stimulated volume and superficial field strength. TMS targets at depths of approximately 4 cm or more results in superficial stimulation strength that may compromise upper limits in TMS safety.

Sekino *et al.* (2015) developed an eccentric figure-of-eight coil that reduced the coil driving current by 20% when compared to the conventional figure-of-eight coil while still inducing similar EF strength.

Koponen, Nieminen and Ilmoniemi (2015) introduced a method to determine the minimum-energy solution for a TMS coil using a spherically symmetric head model for optimisation with given focality constraints. The optimised coil design demonstrated a 73% reduction in power requirement when compared to the figure-of-eight coil with similar focality.

Lu and Ueno (2015) investigated the conventional figure-of-eight coil working with the Halo coil (i.e. Halo–figure-of-eight assembly (HFA) coil), which was computationally analysed for deep TMS in anatomical head model. The HFA coil improved the penetration depth of the magnetic field more than the figure-of-eight coil. In a subsequent work by the same group (Lu and Ueno, 2015a), a figure-of-eight coil working with the circular coil (Halo–circular assembly (HCA) coil) showed an increase at the expense of reduced focality. Further, Lu and Ueno (2017) extended the comparison by including H- and double cone coils. The simulation results demonstrated that double cone, H-, and HCA coils had deeper penetration depth than in the conventional figure-of-eight coil, at the expense of higher and wider spread of induced EF in superficial cortical regions.

Yamamoto *et al.* (2015) proposed a bowl-shaped coil that induces EFs in a wider area of the brain than a figure-of-eight coil. The electromagnetic characteristics of the coil were analysed. A more uniform induced EF can reduce the burden of coil-positioning error but at the cost of focality.

Guadagnin *et al.* (2016) conducted a comparison of 16 different coils (figure-of-eight, large circular, H1-, double cone coils) for deep TMS. The EF distributions were calculated in several brain structures of a head model. The results showed that only the coils of the double cone family were able to reach the distance of deep brain regions (> 4 cm from the cortex); however, this method demonstrated lower focality.

Rastogi *et al.* (2017) proposed a quadruple butterfly coil (QBC) with a high permeability ferromagnetic material acting as a passive magnetic shield of semi-circular shape. The QBC with a shield was compared with a QBC without a shield and the figure-of-eight coil in 50 anatomically realistic heterogeneous head models targeting two brain regions: the vertex and the dorsolateral prefrontal cortex. The shielding solutions showed an improvement in focality of 20% when compared to the conventional figure-of-eight coil and 12% when compared to QBC alone.

Wei *et al.* (2017) investigated multi-coil array optimisation by investigating induced EF in a spherical head model. Marginal improvement was observed for the multi-coil arrays when compared to the figure-of-eight coil in terms of the half-depth distance.

Koponen *et al.* (2017) developed a TMS coil optimisation method in a realistic head geometry with an arbitrary overall coil shape to increase the energy efficiency for focal stimulation. They used the BEM with three-layer head models for computing the induced EF on the cerebral cortex. The optimisation could increase TMS coil efficiency by a factor of two compared to the standard figure-of-eight coil.

Iwahashi *et al.* (2017) proposed a method to evaluate the average coil performance for a group of individuals. To demonstrate the effectiveness, 10 head models comprised of 10 tissues were used. The results showed that there was no remarkable difference between six coils (figure-of-eight coils with and without shielding, eccentric figure-of-eight-type coils) for selectively inducing the maximum EF within the region of interest, although the focality could be improved by considering metallic plates (passive shielding).

Samoudi *et al.* (2018) proposed the double cone coil with the Halo coil (i.e. Halo–double cone assembly (HDA)) and compared it with the HFA, double cone, and Halo coils. Computational analysis of the induced EFs reaching the hippocampus, nucleus accumbens, and cerebellum in a realistic head model showed that only the HDA coil reached the hippocampus and nucleus accumbens with an EF larger than 50% of the maximum value in the cerebral cortex.

Gomez, Goetz and Peterchev (2018) presented a methodology for optimisation of TMS coils. A multi-objective optimisation technique was used for computationally designing TMS coils that achieved optimal trade-offs between EF focality in spherical and MRI-derived head models.

Wang *et al.* (2018) presented a pipeline to produce spherical-shaped cap coils that can reliably replicate the induced EF distribution on the cortex generated by existing TMS coils while significantly reducing energy. Simulations in a realistic head model demonstrated that the EF induced by the cap coil matched that induced by the original coil in both superficial and deep brain regions.

Gomez-Tames *et al.* (2020) compared TMS coil designs for targeting deep brain regions for 18 subjects. For optimised coil positioning to target deep brain regions, the highest EF generated in deep brain regions was 50% of the maximum value in the cortex for the HCA. The systematic analysis also confirmed the trade-off between spread and penetration, where the double cone type coil demonstrated the best performance.

In summary, computational modelling studies were conducted to improve focality, depth, and power requirements, as listed in Table 2. In general, smaller coils have superior focality but lower depth (Thielscher and Kammer, 2004; Deng, Lisanby and Peterchev, 2013; Sekino *et al.*, 2015), while larger coils favour deeper targets (Roth, Zangen and Hallett, 2002; Deng, Lisanby and Peterchev, 2014; Lu and Ueno, 2017; Samoudi *et al.*, 2018). Nevertheless, all coils are subject to a trade-off between depth and focality (Deng, Lisanby and Peterchev, 2014; Guadagnin *et al.*, 2016; Gomez, Goetz and Peterchev, 2018; Gomez-Tames *et al.*, 2020). Shielding approaches may increase the focality at the expense of a reduction in the maximum EF (Kim, 2006; Hernandez-Garcia, 2010; Iwahashi, 2017); further, multi-channel coils have been investigated to improve focality; however, no significant improvement with respect to the figure-of-eight coil was achieved and the method demonstrated difficulty in practical implementation (Kim, Georghiou and Won, 2006; Lu *et al.*, 2009; Wei *et al.*, 2017). Among standard commercial coils for deep TMS, the double cone coil offers a balance between stimulated volume and superficial field strength (Deng, Lisanby and Peterchev, 2014; Guadagnin *et al.*, 2016; Gomez-Tames *et al.*, 2020). Multi-objective optimisation of the coil windings can reduce the required power and reach the physical limits of the trade-off between depth and spread (Hernandez-Garcia *et al.*, 2010; Koponen, Nieminen and Ilmoniemi, 2015; Koponen *et al.*, 2017; Gomez, Goetz and Peterchev, 2018; Wang *et al.*, 2018). The spherical head model provides a standardised platform to evaluate and compare coil designs but with limitations (refer subsection 3.1). Conversely, systematic evaluation of the coil performance in a group of anatomically realistic head models may present more robust analysis (Iwahashi *et al.*, 2017; Rastogi *et al.*, 2017; Gomez-Tames *et al.*, 2020).

### 3.7 Guiding TMS dose

There are no easily measurable responses for the activation of cortical areas other than the areas related to motor/language/phosphene functions. An initial approach was to use the excitation threshold measured in the motor cortex to estimate the cortical excitability at other cortical sites. In this subsection, six studies have been identified that used computational dosimetry to simplify the selection of stimulation parameters.

Stokes *et al.* (2013) used a realistic head model to show that the coil-cortex distance was approximately linearly proportional to the EF induced in the cortex. They proposed the utilisation of the coil-cortex distance as a correction factor to adjust the TMS intensity for other cortical areas based on the measurements in the motor and visual cortices. Two weaknesses were demonstrated, i.e. the intra-individual differences in cortical targets and the effect of coil orientation, which have a large influence on the stimulation efficiency.

Janssen, Oostendorp and Stegeman (2014) utilised EF calculations in a realistic head model and showed that a simple correction based on the inverse of the coil-cortex distance does not adjust the induced EF for regions other than the motor cortex.

Janssen and Oostendorp (2015) examined the induced EF for different coil orientations in 14 cortical targets of one head model (eight tissues). The EF perpendicular to the anterior sulcal wall of the central sulcus was highly susceptible to coil orientations and had to be adjusted for maximising the EF in the motor cortex. Small orientation changes (10°) did not alter the induced EF drastically. Orienting the TMS coil based on anatomical information (MRI) about the targeted brain area can improve the EF, though those orientations determined in one model may be suboptimal for other individuals.

Opitz *et al.* (2016) proposed a TMS guiding method by targeting the EF in specific brain regions associated with functional network maps based on resting-state functional MRI (fMRI). A simulated atlas of regions with low coil orientation-sensitivity can be provided in the absence of TMS dosimetry and fMRI data to personalise coil parameters.

Gomez-Tames *et al.* (2018) developed an atlas to guide the coil orientation and position to group-level optimisation using 18 head models. A universal optimal coil orientation applicable to most subjects was feasible at the primary somatosensory cortex and primary motor cortex. The optimal coil orientation corresponded to an induced EF direction perpendicular to the sulcus wall following the anatomical shape of the hand motor area. Individualised computation of the induced EF became more important in other cortical regions, which had higher inter-subject variability of the cortical folding.

Li *et al.* (2019) used an optimisation technique to reduce the number of computations to determine the optimal TMS coil configuration to target specific brain regions. Up to 11 iterations of EF computations were required for high accuracy in 13 head models under this test.

In summary, the computation of the induced EF to guide TMS becomes more relevant owing to a lack of easily measurable responses in most of the cortical regions. TMS-induced EF is sensitive to coil orientation that does not allow the application of simplified methods using coil-scalp distance or even simplified head models to estimate the induced EF (Janssen and Oostendorp, 2015). Thus, computation using individualised head models together with TMS coil navigation is the most accurate method to determine the induced EF. Alternatively, a group-level analysis of the induced EF is proposed to guide TMS in a group of subjects (Gomez-Tames *et al.*, 2020).

### 3.8 TMS Localisation and validation

During the application of TMS, the site and size of the stimulated cortical volume are unknown. EF dosimetry combined with electrophysiological measurements can be used to gain insight on the activated neural structures in the brain and to validate the EF models. In this subsection, we identified 11 studies that investigated EF-based metrics for TMS localisation and compared and validated with electrophysiology measurements and direct electrical stimulation (DES) on the cerebral cortex.

*3.8.1 Comparison with electrophysiology measurements*

Thielscher and Kammer (2002) reported the first combination of physiological measurements with induced EF modelling. Based on measured threshold stimulator intensities in four subjects, the field distribution on the individual cortical surface was calculated using a spherical head model. The authors proposed the most likely stimulation point at which the variance of the induced EF strengths over all stimulation sites was minimal (lateral part of the hand knob, which is an anatomical region of the hand motor cortex).

Opitz *et al.* (2013) measured the MEP during TMS targeting the right first dorsal interosseous (FDI) muscle and modelled the induced EF in four subjects. The MEP was measured using two different coil orientations (45° and 90° to the midline) at 25 different locations (5 × 5 grid, 1 cm spacing) over the left motor cortex. There were strong correlations for the regression between MEP amplitudes and the calculated mean EF induced in the M1 ($0.70 < r < 0.91$, $n = 4$). Furthermore, the locations of the highest EF strengths were consistent with blood oxygen level-dependent fMRI measurements while subjects voluntarily moved their right index finger.

Krieg *et al.* (2015) investigated the relationship between induced EFs and cortical activation measured indirectly through functional imaging concurrent with TMS. They observed that decomposing the EF into orthogonal vector components based on the cortical surface geometry (and hence, cortical neuron directions) resulted in significant differences between the regions of the cortex that were active and non-active. Later, Arabkheradmand *et al.* (2019) developed an algorithm based on EF calculations and functional neuronal models for predicting the physiological responses evoked by TMS.

Bungert *et al.* (2017) used MRI-based head models for individualised estimation of the EF induced in nine subjects. The motor thresholds in the FDI and abductor digiti minimi muscles were measured in the same subjects. The authors compared the normal component and strength of the EF with the variations in the measured motor thresholds of two muscles when the coil was rotated. They observed that the EF strength on the crown of the precentral gyrus was significantly related to the measured motor threshold, which indicated that TMS activated a focal region around the gyral crown.

Laakso *et al.* (2018) modelled the motor cortical TMS in MRI-based models of 19 individuals. The AMT and RMT of the FDI muscle were measured at 3 to 5 coil locations. The authors showed that the induced EF in a small region in the 'hand knob' of M1 was significantly related to the measured MTs. At the group-level, the EF in the ventral and lateral part of the hand knob demonstrated approximately 70% variability in the MT owing to coil location.

Mikkonen *et al.* (2018) measured the RMT in the FDI muscle and calculated the induced EF in 28 subjects. The individually calculated mean EF strength in the motor cortex significantly correlated with the measured RMT ($R^2 = 0.44$).

Weise *et al.* (2020) performed motor-cortical-based TMS measurements using several coil locations and orientations in 15 subjects and modelled the induced EFs using MRI-based head models. By investigating the congruence of the calculated EF and the measured MEP amplitudes, the authors showed that the origin of MEPs was around the gyral crowns and upper parts of the sulcal wall, and that the EF strength was the most relevant quantity to explain the observed effects. For validation, the authors optimised the position and orientation of the TMS coil to produce the maximum EF strength at the identified cortical location. The optimised scenario showed a reduction of the TMS intensity to generate similar MEPs, thereby validating the computational model.

Reijonen *et al.* (2020) measured the RMT of the FDI muscle and modelled the induced EF in 10 subjects. The relationship between the calculated EF strength and the measured RMT suggested that the activation site of TMS was focal and located in the hand knob area of the motor cortex.

The studies agree on identifying a significantly localised activation site in the somatotopically organised motor cortex (Krieg *et al.*, 2015; Bungert *et al.*, 2017; Laakso *et al.*, 2018; Weise *et al.*, 2020). However, there is no consensus on the best EF-based metrics (e.g. EF strength or the normal or tangential EF component) or the specific gyral activation site (i.e. crown or upper parts of sulcal wall).

*3.8.2 Comparison with direct electric simulation*

Opitz *et al.* (2014) compared the computationally predicted stimulation area in TMS with the DES in six patients with tumours near precentral regions. The authors used an MEP mapping experiment combined with realistic individual simulations of the EF distribution during TMS. The stimulation areas in TMS and DES showed an overlap of up to 80%. The Euclidean distance between the centre of gravity of the TMS map and that of the DES map was 6 mm and 9 mm, respectively.

Aonuma *et al.* (2018) proposed a post-processing method to determine TMS activation sites by combining the individualised computed EFs for the coil orientations and positions that delivered high MEPs during peritumoral mapping. Peritumoral mapping by TMS was conducted on patients who had intra-axial brain neoplasms located within or close to the motor speech area. The hand motor areas estimated by this proposal and DES were in good agreement (5 mm distance error) in the ipsilateral hemisphere of four glioma patients. The hotspots predicted by the method used by the authors were better than those identified by a navigation system that is based on spherical model computations.

Seynaeve *et al.* (2019) investigated preoperative mapping based on TMS-induced EF computation in 12 patients. By comparing with DES, the authors argued that the weighted average of the induced EFs calculated with a realistic head model demonstrated superior performance in comparison with other metrics (nearest or perpendicular projection from the coil and location of maximum EF strength). The Euclidean distance between TMS estimation and DES mapping was 11 mm.

Comparison with DES showed that functional localization was possible with a prediction error in the order of 5 to 11 mm by TMS dosimetry. One caveat for the comparison between DES and TMS is that the two methods differ in terms of the EF direction. In the former, the EF is radial from the electrode, whereas the latter is not limited to the direction that is normal to the cortical surface. Thus, TMS may activate different circuits within the same gyrus, considering that the motor system is topographically organised.

## 4. Models of Neural Activation

Experimental studies have been fundamental in identifying mechanisms that could explain neural responses to magnetic stimulation. However, they predominantly used indirect and non-invasive measurements, such as brain imaging and biomarkers of physiological responses (i.e. neuromuscular, speech arrest, phosphene). Directly monitoring the neuronal response during magnetic stimulation would facilitate the understanding of the effects of TMS; however, only a few *in-vitro* studies exist. Conversely, *in silico* studies of neuronal activation can provide new insights at a cellular level and optimise stimulation parameters that cannot be achieved by *in-vitro* approaches and imaging modalities. We have identified and reviewed 17 papers that have used biophysical-based neuron models for studying the mechanisms of TMS. A summary of the papers is listed in table 3, and a detailed review of the studies is given in the following subsections.

### 4.1 Multi-compartment conductance-based model approach

Early modelling studies provided mathematical formalism of polarisation and activation of simplified neuronal structures. They used infinite cables in length representing unmyelinated and myelinated axons that are required to understand the effects of magnetic stimulation at the level of the peripheral nervous system (Reilly, 1989; Roth and Basser, 1990). In this subsection, simple models for investigating the coupling with TMS-induced EF are revised.

To apply earlier neuronal models to the CNS, Nagarajan, Durand and Warman (1993) focused on the magnetic stimulation of short-length neuronal structures, in which the activation at axon terminals follows the EF instead of its gradient, such as in long structures. Thereafter, Nagarajan and Durand (1996) also clarified that both primary and secondary field components (not just the primary component) contributed to excitation and provided a generalised cable equation to account explicitly for both components. At the same time, the validity of this generalised 1-D cable equation for magnetic stimulation was shown to be valid not only for isolated axons but also for the axons in nerve bundles.

In Kamitani *et al.* (2001), the authors coupled the external field by transforming the induced EF into an equivalent intracellular current that was injected into each segment of the cable. The authors also described methods to deal with the injected current in branching and at the terminals of neural structures that allowed the analysis of multi-compartmental realistic neocortical neurons. A similar approach can be observed in Wu *et al.* (2016).

Wang, Grill and Peterchev (2018) investigated and added mathematical rigour to the validity and implications of the method presented in (Nagarajan and Durand, 1996) to present an alternative coupling approach, termed quasi-potential method, which was applied in other study as well (Goodwin and Butson, 2015). The quasi-potential method essentially allows coupling of the EF induced by magnetic stimulation to neuronal membranes by integrating the longitudinally induced EF along with the branching structure of the neural cable model, as presented in equation (5).

**4.2 Level of morphological representation**

Neurons in the motor cortex present different susceptibilities to the induced EF, which varies according to location of the neuronal elements and their relative orientation to the induced EF, as well as intensity and waveform of the induced EF (Lazzaro *et al.*, 2001; Di Lazzaro *et al.*, 2004). In this subsection, studies that have used neuronal models to investigate the mechanisms of TMS in the motor cortex have been reviewed.

In the study presented by Hyodo and Ueno (1996), the computational simulation suggested that the termination points of nerves or the bent part of an axon are low threshold stimulation sites when magnetically stimulated with a figure-of-eight coil. Further, Nagarajan, Durand and Hsuing-Hsu (1997) observed similar results when investigating excitation sites for different positions of a round and butterfly coil during *in-vitro* magnetic stimulation.

Kamitani *et al.* (2001) investigated the effects of the induced EF in a realistic multi-compartmental model of a layer 5 pyramidal cell model. The magnetic stimulation acted on the dendrites in neocortical neurons. The simulation showed brief burst firing followed by a silent period of duration, which is comparable to the experimental data of single-pulse TMS. Further, the simulation showed that the neurons were readily activated to TMS under background synaptic inputs in agreement with experimental results that showed that TMS effects are evoked with lower intensity during muscle contraction.

Pashut *et al.* (2011) investigated the complex representation of CNS neurons. They argued that the magnetic stimulation of CNS neurons depolarised the soma, leading to the initiation of an action potential in the initial segment of the axon. Here, passive dendrites affected this process primarily as current sinks, not sources. However, the possible inaccurate implementation in the current injection method was speculated in the conclusion of this work (Wang, Grill and Peterchev, 2018).

Wu *et al.* (2016) implemented a multitude of detailed physiological and morphological properties of pyramidal cells. The activation thresholds and sites were computed to various field directions and pulse waveforms. The dependence of the initiation sites on both the orientation and the duration of the stimulus implies that the cellular excitability might represent the result of the competition between various firing-capable axonal components.

Moezzi *et al.*, (2018) proposed a biophysical model of electromyography (EMG) signal generation based on the feed-forward CNS network coupled with a pool of motoneurons. The simulated EMG signals matched experimental EMG recordings in shape and size.

### 4.3 Multiscale models and applications using induced EF in realistic head models

In addition to neural modelling, small geometrical alterations, tissue heterogeneity, and tissue conductivity can alter the field distribution and therefore affect the site of activation (refer subsections 3.1 and 3.3). The path of the nerve fibres, which can be determined using tractography, also affects the patterns of activation (Opitz *et al.*, 2011; Nummenmaa *et al.*, 2014). This subsection reviews studies that considered realistic neuron models driven by TMS-induced EFs that were computed in realistic human head.

Salvador *et al.* (2011) investigated neuronal responses using a simplified cortical sulcus model for TMS with various structures, including pyramidal neurons, interneurons, and association fibres embedded in the grey matter and projecting to white matter, which were considered to be the cause of the generation of evoked motor responses. They identified changes in the stimulation threshold that could be shaped by field orientation (coil orientation), pulse waveform, and diameter of neurons. The outcome was that TMS preferentially activated different sets of axons depending on their orientation with respect to the induced current. For instance, neurons modelling pyramidal neuron tracts were excited in the white matter where they were bent. Conversely, cortical interneurons and axon collaterals were excited at their axonal terminations. Finally, pyramidal association fibres were stimulated either at their axonal termination or at a sharp axonal bend.

Goodwin and Butson (2015) integrated anatomically realistic head models derived from MRI data and detailed models of pyramidal cells. This work allowed the visualisation of activated axons of pyramidal cells within a patch of cortex on a subject-specific basis.

De Geeter *et al.* (2015) used personalised anisotropic head model tissues with realistic neural trajectories of the subject, obtained from tractography, based on diffusion tensor images. An investigation of the impact of tissue anisotropy showed that its contribution was not negligible. In contrast, the model proved to be less sensitive to the uncertainty of the tissue conductivity values.

De Geeter *et al.* (2016) used an anisotropic head model with white matter fibre tracts obtained from the patient. The computed induced EF corresponded to different coil positions during the speech arrest experiment, in which TMS was delivered to Broca's area. The authors computationally determined the tract that was activated when a speech arrest occurred.

Seo *et al.* (2017) incorporated layer 3 and layer 5 pyramidal neurons into realistic head models that considered the intricate folding patterns of the cortex. They observed that the action potentials were predominantly generated at the initial segment of the axon.

Soldati *et al.*, (2018) used TMS experiments, physiological measurements, and individualised MRI-based computer simulations for the determination of brain stimulation thresholds. The combined approach with established biological axon models enabled the extrapolation of the measured thresholds for sinusoidally varying EFs.

Gomez-Tames *et al.*, (2019) investigated stimulation thresholds computing the effects in pyramidal tracts embedded in the cortical folding by independent implementations of neural and induced EF computations.

Aberra *et al.* (2020) used a variety of realistic models of neurons across the cortical layers to quantify the effect of TMS with several combinations of pulse waveforms and current directions on the activation of individual neurons. The intracortical axonal terminations in the superficial gyral crown and lip regions were activated with the lowest TMS intensity. The neural activation was primarily driven by the field strength, rather than the field component that was normal to the cortical surface. Changing the induced current direction caused a shift in the activation site, which may explain the differences in thresholds and latencies of muscle responses observed in experiments.

### 4.4 Summary

As listed in Table 3, there is a trend of increasing complexity in the morphology of the neuron modelling that is used to investigate the activation thresholds of individual neurons. Also, neuronal model embedded in realistic head models permits the computation of neuronal activation using individualised EFs. Further, recent studies have developed network models that may explain the generation of different evoked responses. These approaches show the possibility of combining experimental TMS parameters (coil design, position, and orientation) with subject-specific modelling to quantify the excitation of cortical neurons.

The mentioned multiscale approaches may be applied in improving the specificity of preoperative mapping of brain functions in neurosurgery (De Geeter *et al.*, 2016). Also, knowledge of the TMS mechanisms at cellular levels can help for clinical diagnosis of electrophysiological responses (Moezzi *et al.*, 2018; Aberra *et al.*, 2020). Moreover, multiscale modelling can provide additional scientific rationale to developed safety limits for electromagnetic exposure protection in safety guidelines/standards (IEEE International Committee on Electromagnetic Safety. Technical Committee 95., Institute of Electrical and Electronics Engineers. and IEEE-SA Standards Board., no date; ICNIRP, 2010), as shown by (Soldati *et al.*, 2018; Gomez-Tames *et al.*, 2019).

## 5. Conclusions

TMS is used in the treatment and diagnosis of neurological diseases or conditions, neurosurgery mapping, and as a marker to investigate brain functions. Computational dosimetry techniques have aided in improving the understanding of the TMS-induced EF and how it is affected by anatomical and biophysical parameters. In particular, this review showed that there is a high consensus on the importance of accurate modelling of the complex cortical folding and surrounding CSF for obtaining accurate prediction of the stimulation site.

EF modelling has matured to a point so that individual anatomic models can be efficiently generated from MRI data using modelling pipelines, which allows the construction of an individual model of each subject who participates in experimental studies. Various computational methods can be used for computing the induced EF in anatomical models. Recent studies have shown that all commonly used computational techniques can provide sufficient numerical accuracy for EF calculations (Saturnino, Madsen and Thielscher, 2019; Gomez *et al.*, 2020; Soldati and Laakso, 2020).

EF dosimetry has been extensively applied for the development of magnetic coils to, e.g., improve the focality of the induced EF while reducing the energy consumption. Despite the inability to model the effects of individual anatomy on the induced EF, simpler spherical EF models are sufficient for the optimisation and characterisation of magnetic coils (Deng, Lisanby and Peterchev, 2013). Computation cannot overcome physical limitations, such as the depth focality trade-off that makes it difficult to design coils to target deep brain areas (Deng, Lisanby and Peterchev, 2013; Gomez, Goetz and Peterchev, 2018; Gomez-Tames *et al.*, 2020).

The second application of EF modelling is to guide the selection of TMS parameters for stimulating brain areas that do not produce directly measurable responses. Studies using subject-specific anatomical models have shown that stimulation can be optimised individually or in a group of subjects (Opitz *et al.*, 2016; Gomez-Tames *et al.*, 2018; Li *et al.*, 2019). In future, this may allow personalised stimulation protocols for rehabilitation or therapy. However, these approaches have not yet been tested experimentally.

Analysing the relationship between the EF and electrophysiological data can reveal the sites activated by TMS. Recent studies have focused on the hand area of the motor cortex and have revealed strong correspondence between individually calculated EF strength and measured muscle responses (Opitz *et al.*, 2013a; Bungert *et al.*, 2017; Laakso *et al.*, 2018; Weise *et al.*, 2020). The results allow the determination of the site of activation in the motor cortex; this far, studies suggest that muscle responses evoked by TMS originate from a focal area near the crown of the precentral gyrus. Accurately localising the activation sites is relevant, for instance, in preoperative mapping for planning tumour resection.

In addition to the above-mentioned relationship between the EF and electrophysiological responses, the validity of EF dosimetry models is supported by EF measurements in experimental phantoms (e.g. (Salinas, Lancaster and Fox, 2009; Nieminen, Koponen and Ilmoniemi, 2015)) and comparison with direct electrical stimulation (Opitz *et al.*, 2014; Aonuma *et al.*, 2018; Seynaeve *et al.*, 2019) or neuroimaging (Opitz *et al.*, 2013b; Ottenhausen *et al.*, 2015; Arabkheradmand *et al.*, 2019). Validation and verification of the computed induced EF using *in-vivo* and *ex-vivo* measurements in humans can help to tuned further biophysical parameters to have more accurate predictions (Li *et al.*, 2017; Opitz *et al.*, 2017; Vöröslakos *et al.*, 2018).

Combining EF dosimetry with neuron models, i.e., multiscale modelling, can provide a deeper understanding of the neural mechanisms of TMS. State-of-art models can consider morphologically realistic neuron models embedded in individualised head models (Goodwin and Butson, 2015; Seo *et al.*, 2017; Aberra *et al.*, 2020). For instance, multiscale models can reveal the types and locations of activated neurons, and they can also be used to study the effects of pulse waveform and EF direction. While such models can explain many characteristics of evoked responses, the model predictions have not yet been fully validated in experiments. Future studies can that combine multiscale models and experimentally measured responses are needed.

Despite many research uses, EF dosimetry in realistic models is not yet a part of clinical workflow. Recent technological progress has been made towards using EF dosimetry in clinical applications. Progress has been made in automatic generation of head models from MRI data (Rashed, Gomez-Tames and Hirata, 2019, 2020b; Sendra-Balcells *et al.*, 2020) and approaches for computation of EF in real time have been developed (Laakso and Hirata, 2012; Stenroos and Koponen, 2019; Yokota *et al.*, 2019). These advances can allow integration of EF dosimetry as a part of existing neuronavigation systems, which currently employ spherical models for EF estimation. The added value for clinical applications would come from improved accuracy of neuronavigation, e.g., for preoperative planning. Also, stimulation atlas can be derived for specific populations when time- and cost- constraints exist in resources in small clinics and even hospitals due to operation time limitations.

   Research using multiscale modelling can provide a better understanding of the types and locations of activated neurons, which can potentially enable new TMS-based biomarkers for neurological diseases. Further, this could lead to new ways to optimise stimulation to activate a desired set of neurons, which could improve the value of TMS in treatment and rehabilitation.

# 6. Acknowledgement

This work was supported by JSPS Grant-in-Aid for Scientific Research (No. 17H00869 and No. 19K20668) from the Japanese Society for the Promotion of Science (JSPS) and grant no. 325326 from Academy of Finland

# Appendix A. Search strategy

A search strategy was developed to retrieved papers for each subsection of section 3. Another search strategy was developed for section 4. The search database was Web of Science covering the time period from 1990 to 10.02.2020. Google scholar engine was used for identifying studies from 2020 that have not been indexed yet in Web of Science. The detail of the search strategy is presented in Table A1.

# FIGURES AND TABLES

**Table 1.** Head model representation by canonical or simplified geometries in various studies

**Table 2.** Metrics for transcranial magnetic simulation coil design and optimisation

**Table 3.** Multiscale studies for magnetic exposure on central nervous system

**Table A1.** Search strategy used to retrieve papers for the different subsections covering time period: 1990-10.02.2020

**Figure 1.** Pipeline example of the generation of individualised head models from magnetic resonance images

**Figure 2.** Electromagnetic computation pipeline. The TMS coil design is based on a realistic coil and the localisation and position can be retrieved from a neuro-navigation system when investigated together with neurophysiological measurements such as motor-evoked potential (MEP). The volume conductor is obtained by assigning the tissues conductivity to the digital head model from the pipeline in Fig. 1. Finally, the numerical computation yields the induced electric field (EF) in the brain

**Figure 3.** Multiscale modelling: (a) Transcranial magnetic simulation (TMS)-induced EF drives neural activation; (b) TMS acts stronger on the neurons in the neocortex. Neurons are arranged in horizontal layers with different cell types and neural connections that can project to other areas of the brain regions (e.g., connection between interneurons IN and pyramidal neurons PN) or spinal cord; (c) Cable equation that is coupled with the TMS-induced EF that represents a myelinated axon. The structure can be extended to consider more complicated morphologies such as bifurcations in the dendrites

**Figure 4.** Historical trend in TMS in three tracks: EF computation, neural modelling, and TMS technology. Tracks 1 and 2 have been combined in the multiscale analysis of TMS

**Figure 5.** Anatomical head models in TMS: (a) Evolution of head model representation complexity; (b) Number of subjects according to age and gender; (c) Number of studies based on target brain region

Note 1: The average age was used when is the only reported

Note 2: Certain studies may have reported either the age or the gender, but not both. In that case, only the information available is included

**Figure 6.** Illustration of TMS-induced EF in spherical and anatomical head models using figure-of-eight coil. The EF is shown in the brain cortex.

**Figure 7.** Box plot distribution of the conductivity values of the most common tissues used across different TMS dosimetry studies

**Figure 8.** Illustration of common TMS coils. From left to right, circular, figure-of-eight, H, and double cone coils

**Table 1**

| Author | Characteristics | | |
|---|---|---|---|
| (a) Canonical | | | |
| Roth, Zangen, and Hallett (2002) | Homogeneous sphere (7 cm) | | |
| Miranda, Hallett, and Basser (2003) | Heterogeneous sphere (4.6 cm): CSF and other | | |
| Thielscher and Kammer (2004) | Homogeneous sphere (8 cm) | | |
| Salinas, Lancaster and Fox (2009) | Heterogeneous sphere (10 cm, 1 and 4 layers) | | |
| Hernandez-Garcia *et al.* (2010) | Homogeneous sphere (7.5 cm) | | |
| Deng, Lisanby and Peterchev (2013, 2014) | Homogeneous sphere (8.5 cm) | | |
| Nummenmaa *et al.* (2013) | Homogeneous sphere (globally best-fitted to inner-skull surface) Homogeneous sphere (locally fitted to inner-skull surface close to TMS coil location) | | |
| Koponen, Nieminen and Ilmoniemi (2015) | Homogeneous sphere (8.5 cm) | | |
| Yamamoto *et al.* (2016) | Homogeneous sphere (7.5 cm) | | |
| Wei *et al.* (2017) | Homogeneous sphere (8.5 cm) | | |
| (b) Simplified | Characteristics | Acquired Method | No Subjects |
| Thielscher and Kammer (2002) | Sphere manually fitted to the inner surface of the skull | 1.5T MRI (T1) | 4 subjects (25-38 y.o., one female) |
| Kim, Georghiou and Won (2006) | Norman model (homogeneous) | N.A | 1 male (age N.A) |
| Silva, Basser and Miranda (2008) | Idealised gyrus-sulcus | N.A | N.A |
| Salvador *et al.* (2009) | Head-shaped Homogeneous | MRI | 1 Subject (gender and age N.A) |
| Stokes *et al.* (2013) | Head-shaped Homogeneous | Phantom (IEEE 1528-2003) | N.A |
| Yamamoto *et al.* (2015) | Anatomical Brain | Phantom (NICT) | 1 male, 22 y.o. |
| Sekino *et al.* (2015) | Anatomical Brain | MRI | 1 (gender and age N.A) |

DTI: refers to diffusion tensor imaging

**Table 2.**

| Metric | Quantity | Description | Studies |
|---|---|---|---|
| Depth | EF-decay | EF vs penetration distance in the brain | (Roth, Zangen and Hallett, 2002; Kim, Georghiou and Won, 2006; Salvador *et al.*, 2009; Hernandez-Garcia *et al.*, 2010; Lu and Ueno, 2015a, 2015b; Sekino *et al.*, 2015; Wei *et al.*, 2017) |
| | Depth $d_{1/x}$ | Depth where EF is larger than $E_{max}/x$ along the line between $E_{max}$ position and the center of the brain[a,b,c] | (Deng, Lisanby and Peterchev, 2013; Guadagnin *et al.*, 2016; Gomez, Goetz and Peterchev, 2018; Gomez-Tames *et al.*, 2020) |
| Focality | Area ($A_{1/x}$) | Cortical area where EF is larger than $E_{max}/x$ | (Im and Lee, 2006; Salvador *et al.*, 2009; Koponen, Nieminen and Ilmoniemi, 2015; Yamamoto *et al.*, 2015; Rastogi *et al.*, 2017) |
| | Volume ($V_\Omega$) | Mean value of EF over domain $\Omega$ | (Hernandez-Garcia *et al.*, 2010) |
| | Volume ($V_{1/x}$) | Volume where EF > $E_{max}/x$ | (Guadagnin *et al.*, 2016; Rastogi *et al.*, 2017; Samoudi *et al.*, 2018) |
| | Volume ($V_{th}$) | Volume where EF > threshold value *th*. Usually normalised by brain volume | (Deng, Lisanby and Peterchev, 2014; Lu and Ueno, 2015a, 2017; Wei *et al.*, 2017; Gomez, Goetz and Peterchev, 2018) |
| | Spread ($S_{1/x}$) | $S_{1/x} = V_{1/x}/d_{1/x}$ | (Deng, Lisanby and Peterchev, 2013; Gomez, Goetz and Peterchev, 2018; Gomez-Tames *et al.*, 2020) |
| Energy | Coil energy | Minimum coil magnetic field energy | (Koponen *et al.*, 2017; Wang *et al.*, 2018) |

[a]$E_{max}$ is usually at the cortex

[b]Variable *x* is usually ½ or $\sqrt{2}$.

[c]Center of the brain was considered under Cz at a height of T3 and T4 (10-20 EEG system) in anatomical head model or centre of spherical head model.

**Table 3.**

| Study | Neural Morphology[a] | Neuronal Elements[b] | Activation Site | Head model[c] | Others |
|---|---|---|---|---|---|
| Nagarajan, Durand and Warman (1993) | Simple | Small axon, GC | Terminals | × | × |
| Nagarajan and Durand (1996) | Simple | Myelinated Axon | N.A | × | × |
| Hyodo and Ueno (1996) | Simple | Myelinated Axon | Terminals/Bending | × | × |
| Nagarajan, Durand and Hsuing-Hsu (1997) | Simple | Myelinated Axon | Terminals/along axon | × | × |
| Kamitani *et al.* (2001) | Realistic | Layer 3 (L3) PN | Dendrites | × | × |
| Pashut *et al.* (2011) | Realistic | L3 PN | Soma | × | × |
| Salvador *et al.* (2011) | Realistic | L5 PN, IN, AF | ◦Fiber bends (PN track) ◦Axonal terminations (interneurons and collaterals ◦Combination (association fibers) | △ | × |
| De Geeter *et al.* (2015) | Simple | PNT, AF | Stimulation tract's position according to TMS coil orientation | ○ | DTI |
| Goodwin and Butson (2015) | Realistic | L3 PN | Initiation at neural elements (dendrite, soma, axon) depends on the coil orientation | ○ | × |
| Wu *et al.* (2016) | Realistic | PN, IN. | Competition of various neuronal elements. Determined by the local geometry and field orientation/waveform | × | × |
| De Geeter *et al.* (2016) | Realistic | PN | No discussed | ○ | DTI Navigation System |
| Seo *et al.* (2017) | Realistic | L3 and L5 PNs | Mostly at axon initial segment and a few near boundary GM/WM | ○ | × |
| Moezzi *et al.* (2018) | Complex | IN synapse onto L5 PN that synapse onto motor neurons | No discussed | × | × |
| Wang, Grill and Peterchev (2018). | Simple | Myelinated Axon | Axonal undulation can affect thresholds | × | × |
| Soldati et al., (2018) | Simple | PNT | Axonal termination in gyrus/lip of crown | ○ | Navigation System |
| Aberra *et al.* (2020) | Realistic | L1 to L4 including Neurogliaform, PN, large basket | Mixed | ○ | × |
| Gomez-Tames *et al.* (2019) | Simple | PNT | Bends for PN tract | ○ | × |

[a] Simple refers to a neuron without bifurcations

[a] PN: pyramidal neuron; PNT: pyramidal neuron track; IN: cortical interneuron; GC: granule cell; AF: associate fibres; AC: axonal collaterals

[c] ○: head models with at least five tissues (scalp/skin, skull, CSF, grey matter, and white matter) with realistic cortical folding representation; △: includes gyral/sulcus structure; ×: otherwise

# Table A1

| 3.1. Representation of head tissues | | | | | | |
|---|---|---|---|---|---|---|
| Search data | TI=(TMS OR rTMS OR (Transcranial OR brain) Magnetic Stimulation) AND TS=("Electric Field$" OR "Volume Conductor$" OR "induced current density") AND TS=(("Head$ " OR Anatomic* OR Brain OR Cortical OR Spher*) NEAR/5 (Comput* OR Model* OR Simulation$ OR Biophysical)) AND TS=(primary field OR secondary field OR displacement current$ OR boundar* $ OR Inhomogene* OR heterogeneit*) | | | | | |
| Identified from database | 38 | Excluded (not relevant) | 31 | Identified from other sources | 0 | Relevant | 7 |
| Included in analysis | 7 | | | | | |
| **3.2 Electrical conductivity: variability** | | | | | | |
| Search data | TI=(TMS OR rTMS OR (Transcranial OR brain) Magnetic Stimulation) AND TS=("Electric Field$" OR "Volume Conductor$" OR "induced current density") AND TS=(("Head$ " OR Anatomic* OR Brain OR Cortical OR Spher*) NEAR/5 (Comput* OR Model* OR Simulation$ OR Biophysical)) | | | | | |
| Identified from database | 104 | Excluded (not relevant) | 38 | Identified from other sources | 0 | Relevant | 66 |
| Included in analysis | 66 | | | | | |
| **3.3. Electrical conductivity: anisotropy** | | | | | | |
| Search data | TI=(TMS OR rTMS OR (Transcranial OR brain) Magnetic Stimulation) AND TS=("Electric Field$" OR "Volume Conductor$" OR "induced current density") AND TS=("Head$ " OR Anatomic* OR Brain OR Cortical OR Spher*) AND TS=(Anisotropy) | | | | | |
| Identified from database | 18 | Excluded (not relevant) | 14 | Identified from other sources | 0 | Relevant | 4 |
| Included in analysis | 4 | | | | | |
| **3.4. TMS coil models and verification** | | | | | | |
| Search data | TI=(TMS OR rTMS OR (Transcranial OR brain) Magnetic Stimulation) AND TS=("Electric Field$" OR "Volume Conductor$" OR "induced current density") AND TS=("Head$ " OR Anatomic* OR Brain OR Cortical OR Spher*) AND TS=("coil model*" OR "coil wir*" OR (measure* OR validat* OR accura*) NEAR/4 (coil$)) | | | | | |
| Identified from database | 24 | Excluded (not relevant) | 17 | Identified from other sources | 0 | Relevant | 7 |
| Included in analysis | 7 | | | | | |
| **3.5 Effects of anatomical and inter-individual factors** | | | | | | |
| Search data | TI=(TMS OR rTMS OR (Transcranial OR brain) Magnetic Stimulation) AND TS=("Electric Field$" OR "Volume Conductor" ) AND TS=(("Head$ " OR Anatomic* OR Brain OR Cortical) NEAR/5 (Comput* OR Model* OR Simulation$ OR Biophysical)) AND TS=((gyrus OR gyral OR sulcus OR sulci OR variability OR individual$ OR subject$)) | | | | | |
| Identified from database | 52 | Excluded (not relevant) | 44 | Identified from other sources | 1 | Relevant | 9 |
| Included in analysis | 10 | | | | | |
| **3.6 Coil design: optimisation and performance** | | | | | | |
| Search data | TI=(TMS OR rTMS OR (Transcranial OR brain) Magnetic Stimulation) AND TS=("Electric Field$" OR "Volume Conductor" ) AND TS=(("Head$ " OR Anatomic* OR Brain OR Cortical) NEAR/5 (Comput* OR Model* OR Simulation$ OR Biophysical)) AND TS=(coil AND (design OR optimization OR performance)) | | | | | |
| Identified from database | 48 | Excluded (not relevant) | 26 | Identified from other sources | 1 | Relevant | 22 |
| Included in analysis | 23 | | | | | |

**3.7 Guiding TMS dose**

| | | | | | | |
|---|---|---|---|---|---|---|
| Search data | TI=(TMS OR rTMS OR (Transcranial OR brain) Magnetic Stimulation)<br>AND TS=("Electric Field$" OR "Volume Conductor" )<br>AND TS=(("Head$ " OR Anatomic* OR Brain OR Cortical) NEAR/5 (Comput* OR Model* OR Simulation$  OR Biophysical))<br>AND TS=(guide OR atlas OR  target* OR coil-target distance ) | | | | | |
| Identified from database | 39 | Excluded (not relevant) | 34 | Identified from other sources | 1 | Relevant | 5 |
| Included in analysis | 6 | | | | | |

**3.8 Localising TMS**

| | | | | | | |
|---|---|---|---|---|---|---|
| Search data | TI=(TMS OR rTMS OR "Transcranial Magnetic Stimulation")<br>AND TS=("Electric Field$" OR "Volume Conductor" )<br>AND TS=(("Head$ " OR Anatomic* OR Brain OR Cortical) NEAR/5 (Comput* OR Model* OR Simulation$  OR Biophysical))<br>AND TS=((physiologic* OR Electrophysiolog*) AND measureme* OR MEP OR fMRI OR PET OR DES OR "motor threshold$" ) | | | | | |
| Identified from database | 27 | Excluded (not relevant) | 16 | Identified from other sources | 1 | Relevant | 11 |
| Included in analysis | 12 | | | | | |

**4.  Nerve modelling**

| | | | | | | |
|---|---|---|---|---|---|---|
| Search data | TI=(TMS OR "Magnetic Stimulation" OR electromagnetic OR "Induced Electric Field$" OR "Magnetic Field Stimulation")<br>AND TS=("Electric Field$" OR "Volume Conductor")<br>AND TS=(Brain$ OR Cortex OR Head$)<br>AND TS=(("I-wave$" OR  "D-wave$" OR "Neuron*" OR "Interneuron" or Axon$ OR Nerve$ OR "pyramidal" OR "White Matter") NEAR/10 (Comput* OR Multiscale OR Model* OR Simulation$  OR Biophysical*)) | | | | | |
| Identified from database | 42 | Excluded (not relevant) | 27 | Identified from other sources | 3 | Relevant | 15 |
| Included in analysis | 18 | | | | | |

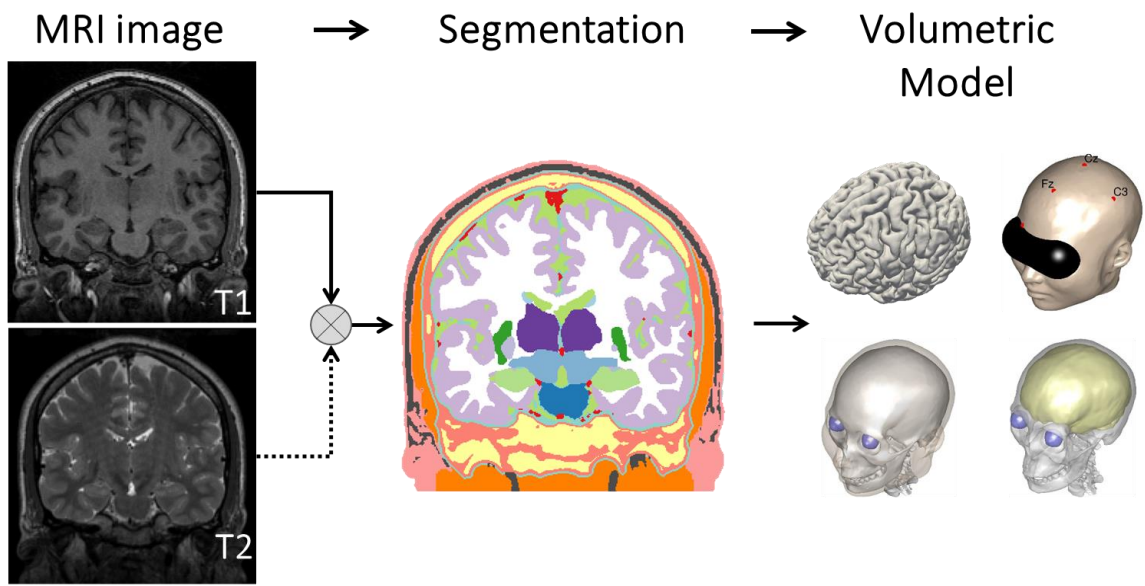

**Figure 1.**

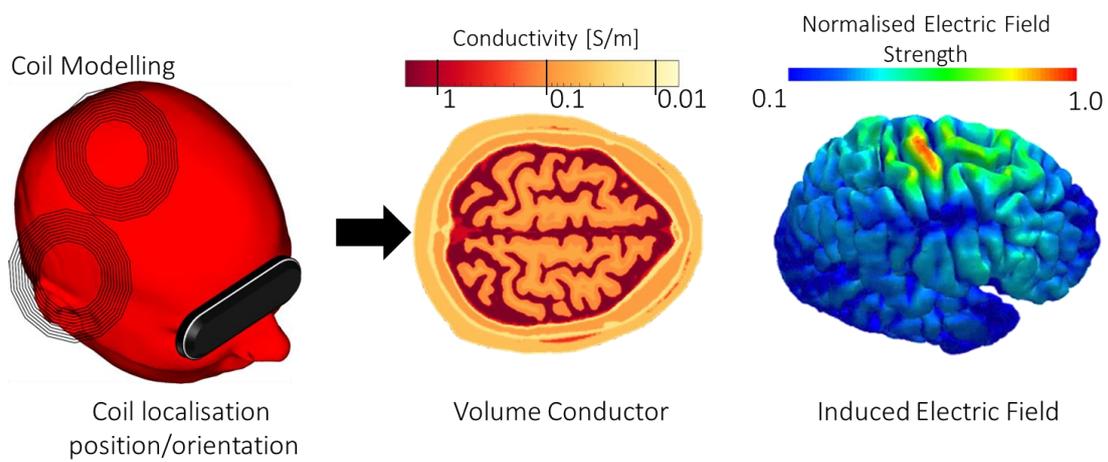

**Figure 2.**

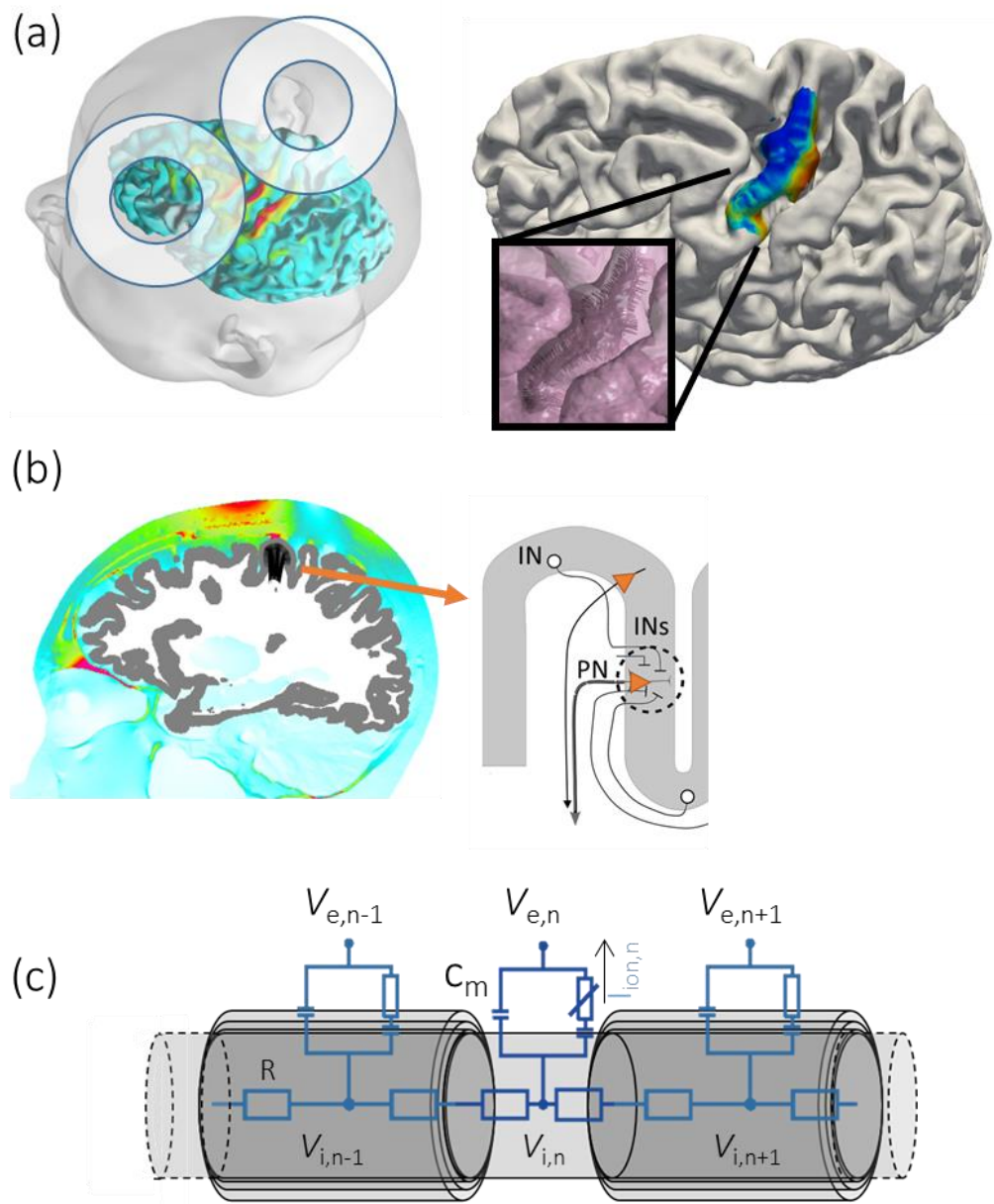

**Figure 3.**

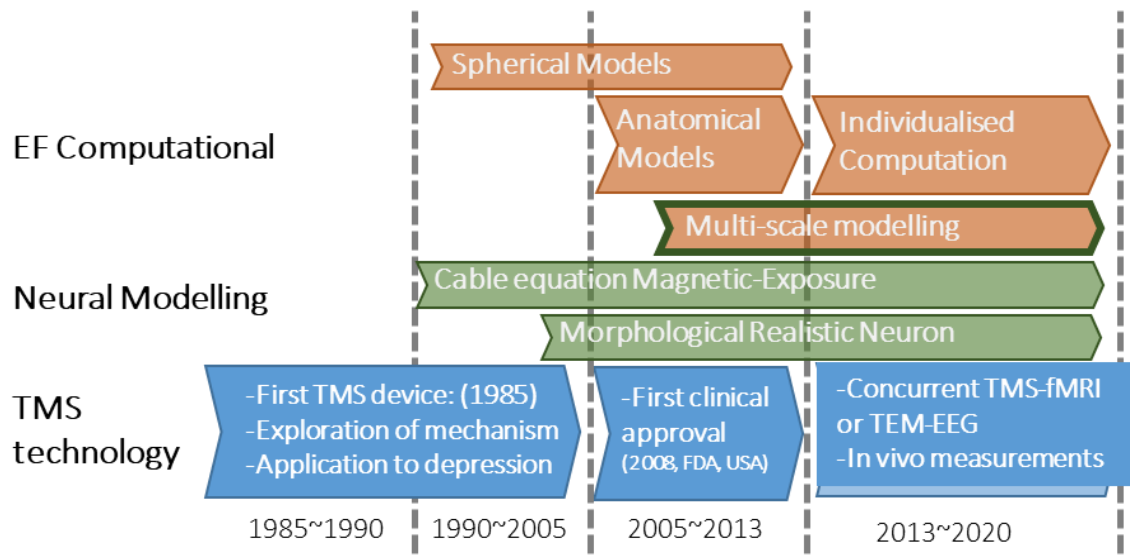

**Figure 4.**

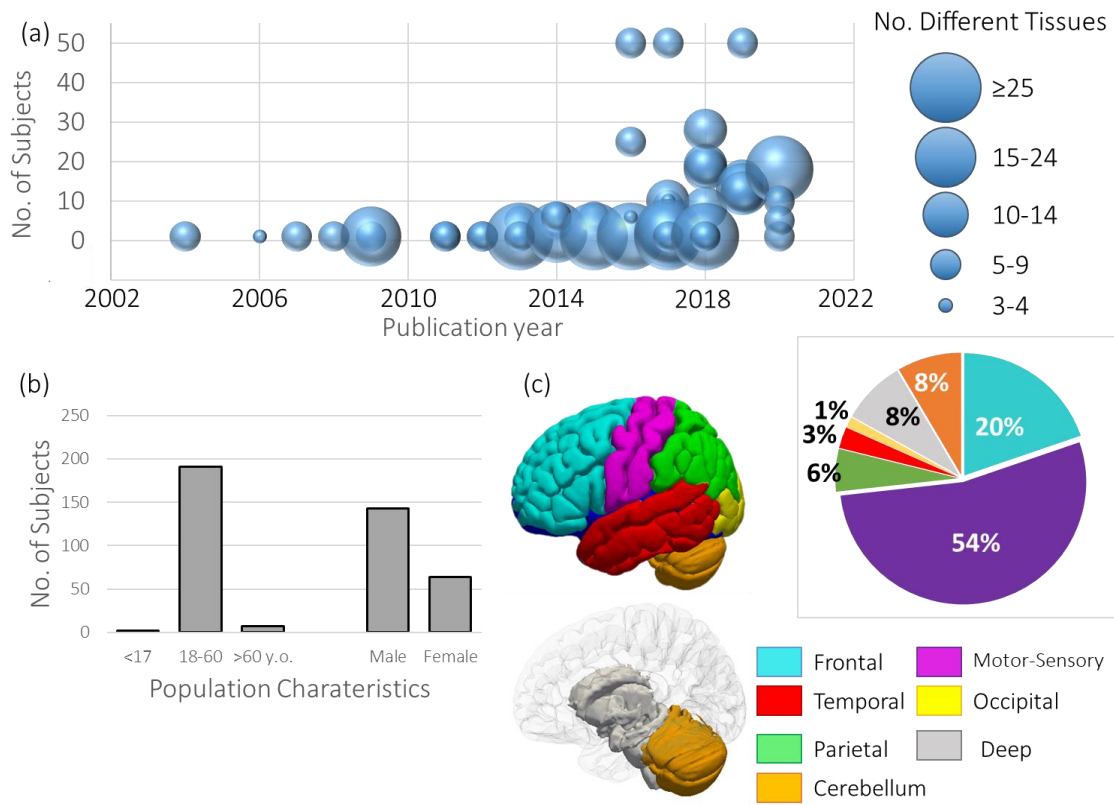

**Figure 5**

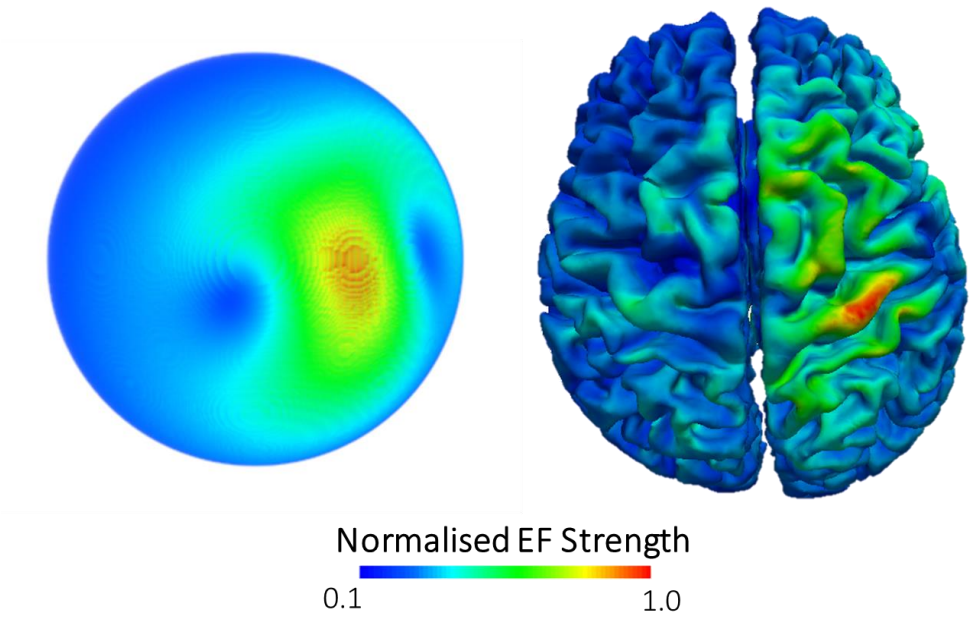

**Figure 6.**

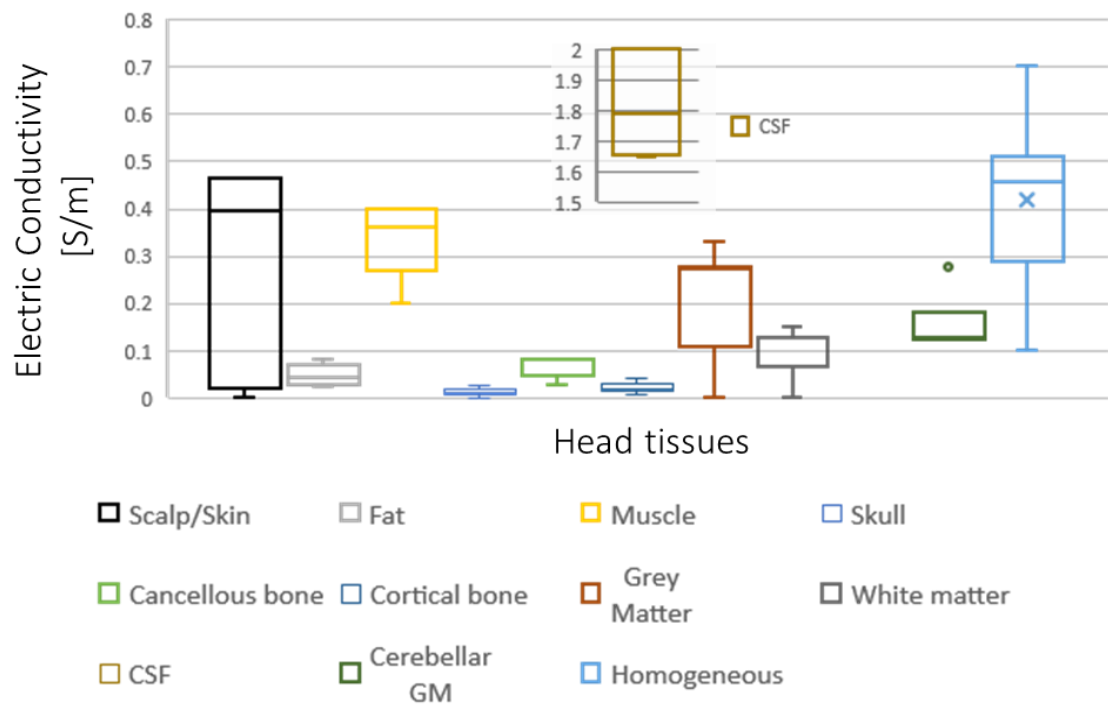

**Figure 7.**

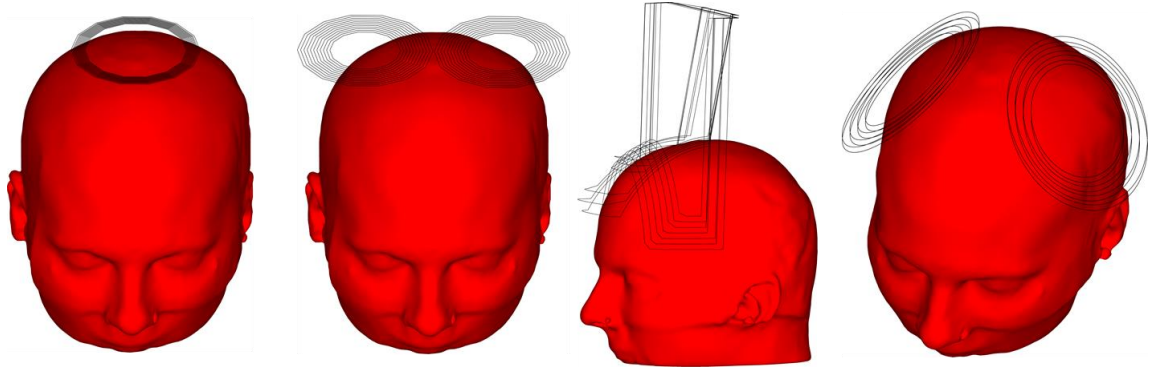

**Figure 8.**